# Cubic Hall viscosity in three-dimensional topological semimetals

Iñigo Robredo,[1,2,*] Pranav Rao,[3,*] Fernando de Juan,[1,4] Aitor Bergara,[1,2,5] Juan L. Mañes,[2] Alberto Cortijo,[6,7] M. G. Vergniory,[1,4,†] and Barry Bradlyn[3,‡]

[1]*Donostia International Physics Center, 20018 Donostia-San Sebastian, Spain*
[2]*Department of Physics, University of the Basque Country UPV/EHU, Apartado 644, 48080 Bilbao, Spain*
[3]*Department of Physics and Institute for Condensed Matter Theory, University of Illinois at Urbana-Champaign, Urbana, Illinois 61801-3080, USA*
[4]*IKERBASQUE, Basque Foundation for Science, Maria Diaz de Haro 3, 48013 Bilbao, Spain*
[5]*Centro de Física de Materiales CFM, CSIC-UPV/EHU, Paseo Manuel de Lardizabal 5, 20018 Donostia, Basque Country, Spain*
[6]*Departamento de Física de la Materia Condensada, Universidad Autónoma de Madrid, Madrid E-28049, Spain*
[7]*Condensed Matter Physics Center (IFIMAC), Madrid E-28049, Spain*

The nondissipative (Hall) viscosity is known to play an interesting role in two-dimensional (2D) topological states of matter, in the hydrodynamic regime of correlated materials, and in classical active fluids with broken time-reversal symmetry (TRS). However, generalizations of these effects to 3D have remained elusive. In this work, we address this question by studying the Hall viscoelastic response of 3D crystals. We show that for systems with tetrahedral symmetries, there exist new, intrinsically 3D Hall viscosity coefficients that cannot be obtained via a reduction to a quasi-2D system. To study these coefficients, we specialize to a theoretically and experimentally motivated tight-binding model for a chiral magnetic metal in (magnetic) space group [(M)SG] $P2_13$ (No. 198.9), a nonpolar group of recent experimental interest that hosts both chiral magnets and topological semimetals (TSMs). Using the Kubo formula for viscosity, we compute two forms of the Hall viscosity, phonon and "momentum" (conventional) and show that for the tight-binding model we consider, both forms realize the novel cubic Hall viscosity. We conclude by discussing the implication of our results for transport in 2D magnetic metals and discuss some candidate materials in which these effects may be observed.

*Introduction.* The discovery of hydrodynamic flow in two-dimensional (2D) metallic systems [1,2] has renewed interest in the study of nondissipative "Hall" viscosity. In rotationally invariant 2D fluids, there is a single Hall viscosity coefficient related to the topological properties of the occupied electronic states [3–11]. In clean systems, the Hall viscosity manifests in width-dependent corrections to the Hall conductance of mesoscopic channels, backflow corrections to the local current density near point contacts, and in moments of the semiclassical distribution function [12–15]. Local voltage measurements on graphene samples in magnetic fields have shown signatures of the Hall viscosity [16]. Hall viscosity also appears in classical fluids with broken time-reversal symmetry (TRS) like chiral active fluids [17,18]. This "momentum" Hall viscosity (MHV) describes a stress response that can be related to a change in momentum density. The MHV contains meaningful information even beyond the hydrodynamic regime [3,19–21].

In parallel, a related geometric response coefficient— the phonon Hall viscosity (PHV)—has gained attention. A response to dynamic strains via electron-phonon coupling, and also a rank four tensor, the PHV is expected to appear in the dispersion for acoustic phonons [20,22] and in spin-phonon-coupled systems through a contribution to thermal Hall conductance [23–26].

Beyond 2D, the role of nondissipative viscosity in transport remains largely unexplored. Reports of hydrodynamic behavior in topological semimetals (TSMs) [27], and the growing interest in magnetic TSMs [28], raise the question of how to generalize the Hall viscosity to 3D. Preliminary efforts have focused on quasi-2D transport [29–35], or made use of preferred "polar" directions such as the Weyl node separation direction in TSMs. Furthermore, octahedral symmetry forbids the presence of a nonzero Hall viscosity [19,36]. However, magnetic crystals may have nonpolar point group symmetries that are not octahedral; the nondissipative geometric response of such systems remains an open question.

Looking beyond 2D, in this work we find that tetrahedral symmetry allows for the appearance of a new, fundamentally 3D "cubic" Hall viscosity. To our knowledge this has not been encountered before in the literature and could be realized in a wide array of classical and quantum fluids with broken TRS.

---

*These authors contributed equally to this work.
†maiagvergniory@dipc.org
‡bbradlyn@illinois.edu

For uniaxial flows, this new viscosity gives a force perpendicular to the flow direction which vanishes when the velocity is constant along the direction of flow. As a proof-of-principle, we focus on a toy model in the experimentally interesting case of the cubic MSG $P2_13$ (No. 198.9), with TRS breaking chiral magnetism. Chiral multifold fermions such as these act as point sources of Berry curvature in the Brillouin zone [37–45], making them ideal models to explore topological response functions [40,46–48]. We compute the MHV and PHV for this model. For the PHV, we consider an electron-phonon coupling ansatz to derive the "phonon" strain coupling [20,22,34,49,50] which will yield us a resulting "phonon" stress tensor. For the MHV, we use the recently introduced lattice formulation of stress response [11] to derive a coarse-grained strain coupling corresponding to a conserved momentum density. Using the Kubo formula for viscosity [10], we derive both the MHV and the PHV for a spin-1 fermion. We discuss the implication of our work for chiral magnetic TSMs such as the family $Mn_3IrSi$, $Mn_3IrGe$, $Mn_3Ir_{1-y}Co_ySi$, and $Mn_3CoSi_{1-x}Ge_x$ [28,51,52] in MSG $P2_13$ (No. 198.9).

*Hall viscosity with cubic symmetry.* Let us examine the symmetry properties of the Hall viscosity tensor in systems with tetrahedral symmetry. The symmetry analysis in this section holds for both the MHV and the PHV, although the interpretation of the resulting stress differs. The tetrahedral point group (denoted 23) is the simplest cubic group, generated by twofold rotations about the $\hat{\mathbf{x}}, \hat{\mathbf{y}}$, and $\hat{\mathbf{z}}$ axes, as well as a threefold rotation about the $\hat{\mathbf{x}} + \hat{\mathbf{y}} + \hat{\mathbf{z}}$ cubic body diagonal. For details of the irreducible representations (irreps) of the point group 23, we refer the reader to the supplemental material (SM) [53], as well as to the group theory tables on the Bilbao Crystallographic Server [54–56]. We define the Hall viscosity tensor as the antisymmetric (and therefore nondissipative) component of the viscosity tensor $\eta^i{}_j{}^k{}_\ell$ [4,11,19,29,57],

$$(\eta_H)^i{}_j{}^k{}_\ell \equiv \frac{1}{2}\bigl(\eta^i{}_j{}^k{}_\ell - \eta^k{}_\ell{}^i{}_j\bigr), \tag{1}$$

where $i, j, k, \ell$ index the three spatial directions. The Hall viscosity is odd under TRS [58], and for a fluid with a nonuniform velocity field $v^\ell$, leads to a viscous stress [59],

$$\delta\tau^i{}_j = -(\eta_H)^i{}_j{}^k{}_\ell \partial_k v^\ell. \tag{2}$$

Our goal is to identify the independent symmetry-allowed Hall viscosity coefficients. Since these coefficients are scalars, we can determine them by finding all rank four antisymmetric tensors invariant under 23. Introducing the irreducible tensors,

$$\Theta^a_{ij} = \begin{cases} \frac{1}{\sqrt{3}}(\delta_{1i}\delta_{1j} + \delta_{2i}\delta_{2j} - 2\delta_{3i}\delta_{3j}) & a=1 \\ \delta_{1i}\delta_{1j} - \delta_{2i}\delta_{2j} & a=2 \end{cases} \tag{3}$$

$$\Lambda_{ijk} = |\epsilon_{ijk}|, \tag{4}$$

with $\epsilon_{ijk}$ the Levi-Civita symbol, we can form two invariant tensors and thus identify two viscosity coefficients compatible with tetrahedral symmetry:

$$\begin{aligned}(\eta_H)^i{}_j{}^k{}_\ell &= -\eta_1 \epsilon_{ab} \Theta^{ai}{}_j \Theta^{bk}{}_\ell + \frac{\eta_2}{\sqrt{3}}\bigl(\Lambda^{mi}{}_\ell \epsilon^k{}_m{}_j - \Lambda^{mk}{}_\ell \epsilon_m{}^i{}_j\bigr) \\ &= \eta_1(\lambda_3 \wedge \lambda_8)^i{}_j{}^k{}_\ell + \frac{i\eta_2}{\sqrt{3}}(\lambda_1 \wedge \lambda_2 + \lambda_6 \wedge \lambda_7 - \lambda_4 \wedge \lambda_5)^i{}_j{}^k{}_\ell,\end{aligned} \tag{5}$$

where $\epsilon_{ab}$ is the 2D Levi-Civita symbol. In the second line we have reexpressed the antisymmetric product of irreducible tensors in terms of the Gell-Mann matrices $\lambda$ [53]. This shows that the $\eta_2$ term is the antisymmetric dot product of matrices transforming in two 3D T (vector) irreps,

$$\begin{aligned}\boldsymbol{L} &\equiv T(\lambda_7, -\lambda_5, \lambda_2) \\ \tilde{\boldsymbol{L}} &\equiv T(\lambda_6, \lambda_4, \lambda_1),\end{aligned} \tag{6}$$

each spanned by a triplet of Gell-Mann matrices.

Crucially, neither of $\eta_{1,2}$ require a preferred spatial direction. This contrasts with the familiar "quasi-2D" Hall viscosities which are proportional to a pseudovector (i.e., a magnetic field). Thus, $\eta_1$ and $\eta_2$ are new, essentially 3D Hall viscosities, which can be nonzero in systems with broken rotational symmetry. By contrast, octahedral symmetry requires $\eta_1 = \eta_2 = 0$, as the two tensors in Eq. (5) do not transform in the trivial representation of the group 432 ($O$). Furthermore, $\eta_1$ and $\eta_2$ can be nonzero in centrosymmetric point groups such as $T_h$ ($m\bar{3}$). Next, we compute the viscous force density that is produced by these Hall viscosities, noting a difference in interpretation for forces due to MHV and PHV [60],

$$f^\eta_j = -\partial_i \delta\tau^i{}_j = (\eta_H)^i{}_j{}^k{}_\ell \partial_i \partial_k v^\ell, \tag{7}$$

where $\boldsymbol{f}^\eta$ is the force density and $\delta\tau^i{}_j$ is the viscous stress tensor. We find that $\eta_1$ and $\eta_2$ contribute additively to $\boldsymbol{f}^\eta$:

$$f^\eta_j = \frac{\eta_1 + \eta_2}{\sqrt{3}} \Lambda^{mik} \partial_i \partial_k (\epsilon_{mj\ell} v^\ell). \tag{8}$$

We see that the fully symmetric tensor $\Lambda$, which is only invariant in systems with tetrahedral symmetry, plays a key role in generating the nondissipative forces. Contrast this with quasi-2D Hall viscous forces, which take the form,

$$f^{\eta,2D}_j = \eta_{2D} B^m \nabla^2 (\epsilon_{mj\ell} v^\ell), \tag{9}$$

and require a symmetry-breaking pseudovector $\mathbf{B}$.

Finally, since only the sum $\eta_1 + \eta_2$ appears in the viscous forces, there must exist a divergenceless contact term which shifts between $\eta_1$ and $\eta_2$ in the bulk. This term is

$$\delta\tau^i{}_j = C_0 \epsilon^{mik} \Lambda_{mj\ell} \partial_k v^\ell, \tag{10}$$

which shifts

$$\eta_1 \to \eta_1 + \frac{\sqrt{3}C_0}{2} \tag{11}$$

$$\eta_2 \to \eta_2 - \frac{\sqrt{3}C_0}{2}, \tag{12}$$

analogous to the bulk redundancy between Hall viscosity and odd pressure in 2D systems [11,61]. We show the effects of $\eta_{1,2}$ in Fig. 1.

*Tight-binding model.* Let us now consider a model for a cubic chiral magnetic system and compute $\eta_{1,2}$ for both the MHV and the PHV as a proof-of-principle [62]. Our tight-binding Hamiltonian is

$$H = \sum_{nm,\mathbf{r},\mathbf{r}'} c^\dagger_{n\mathbf{r}'} t^{\mathbf{r},\mathbf{r}'}_{nm} c_{m\mathbf{r}}, \tag{13}$$

consisting of $s$-type orbitals at $4a$ Wyckoff position of SG $P2_13$ (No. 198). The indices $n$ and $m$ label the four orbitals,

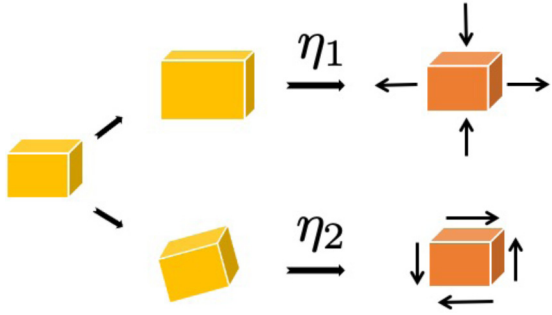

FIG. 1. Schematic of $\eta_1$ and $\eta_2$. Dynamic strains (yellow) and viscosity give rise to stresses (orange). In response to a dynamic strain that elongates the length and width of a cubic parcel of fluid while compressing the depth, $\eta_1$ produces a diagonal shear stress. In response to a dynamic rotation of the parcel, $\eta_2$ produces an off-diagonal shear stress

and $\mathbf{r}$, $\mathbf{r}'$ index the unit cells of the crystal with lattice spacing $a$. The nearest-neighbor hopping $t_{nm}^{\mathbf{r},\mathbf{r}'}$ has uniform magnitude $t$, and we break TRS with a cubic-symmetric magnetic flux $\phi$ via a Peierls substitution [63]. Shifting to momentum space and suppressing the orbital $n$ and $m$ indices, we write

$$H = \sum_{nm\mathbf{k}} c_{\mathbf{k}}^{\dagger} f(\mathbf{k}) c_{\mathbf{k}}. \quad (14)$$

We give $f(\mathbf{k})$ explicitly in the SM [53], and we show the spectrum in Fig. 2(a). As a simple example of where to find nonvanishing cubic Hall viscosity, we focus on the physics near the $\Gamma = (k_x, k_y, k_z) = (0, 0, 0)$ point. Expanding the Hamiltonian around $\Gamma$ and working in the basis of a nondegenerate (spin-0) and a threefold degenerate (spin-1) bands,

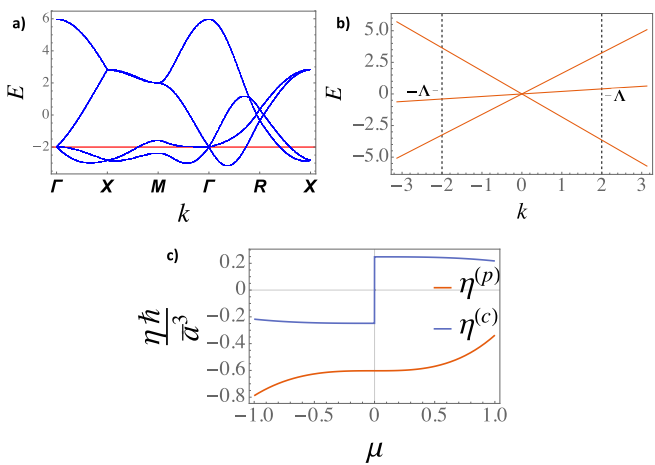

FIG. 2. (a) Band structure of the full tight-binding model Eq. (15). The gap at the $\Gamma$ point scales with the hopping strength $t$. (b) Effective three band description of the $\Gamma$ point for $\mu$ near the threefold degeneracy, with $\Lambda$ a momentum cutoff chosen to regulate the Hall viscosity, which is plotted in (c) for the PHV "(p)" and MHV "(c)" methods (in the plot $\bar{a}^3 = \frac{a^3}{8\pi^3}$).

we find [40]

$$f(\mathbf{k}) \approx \begin{pmatrix} 6t\cos(\phi) & iv_F e^{i\phi} \mathbf{k}^T \\ -iv_F e^{-i\phi} \mathbf{k} & h \end{pmatrix}, \quad (15)$$

where $v_F = ta$. The lower-right block corresponds to the spin-1 bands and can be expressed in terms of the vectors $\mathbf{L}$ and $\tilde{\mathbf{L}}$ as,

$$h = v_F(\cos(\phi)\mathbf{k} \cdot \mathbf{L} + \sin(\phi)\mathbf{k} \cdot \tilde{\mathbf{L}}) - 2t\cos(\phi)\lambda_0, \quad (16)$$

when $\phi = 0$, $h$ describes an $SO(3)$-invariant spin-1 fermion. The vector $\tilde{\mathbf{L}}$ in Eq. (6) parameterizes the breaking of $SO(3)$ to the discrete point group 23.

From Eq. (15), we see that when $\phi$ is small there is a gap of order $t$ separating the spin-0 and spin-1 fermions at $\Gamma$. Thus for small $\phi$ and $\mathbf{k}$, transitions from the threefold to onefold degeneracies mediated by the off-diagonal elements of $f(\mathbf{k})$ are parametrically small, and we can restrict our attention to the spin-1 fermion. Furthermore, in the continuum limit $a \to 0$ with $v_F = ta$ fixed, we see that the gap becomes infinitely large. With this in mind, we focus specifically on the threefold fermion.

*Stress response.* To compute the MHV and PHV, we employ the stress-stress form of the Kubo formula [10] (in the limit $\omega^+ \to 0$),

$$(\eta^H)^\mu{}_\nu{}^\lambda{}_\rho = \frac{1}{2\omega^+V} \int_0^\infty dt e^{i\omega^+ t} (\langle [T_\nu^\mu(t), T_\rho^\lambda(0)] - [T_\rho^\lambda(t), T_\nu^\mu(0)] \rangle). \quad (17)$$

To do so we first must define a stress tensor,

$$T_{\mu\nu} = \sum_{nm\mathbf{k}} c_{n\mathbf{k}}^{\dagger} T_{\mu\nu}(\mathbf{k}) c_{m\mathbf{k}}, \quad (18)$$

corresponding to Eq. (13). For the PHV, we define the stress by considering an electron-phonon coupling ansatz [20,22,64] and perturbing the background lattice, yielding the *phonon* stress tensor. The phonon stress results from microscopically perturbing the lattice via phonons, relating to atomic displacements. For the MHV, we perturb the electronic degrees of freedom directly via coupling to background geometry [10,11], yielding the *continuity* stress tensor. This is a coarse-grained stress tensor that directly corresponds to momentum transport in the long-wavelength limit and can be identified with the stress tensor of fluid dynamics.

In the phonon method, strain is introduced into the model through small displacements of the orbital positions, modifying the hopping parameters $t_{nm}^{\mathbf{r},\mathbf{r}'}$ as

$$t^{\mathbf{r},\mathbf{r}'} \to e^{-(\delta \mathbf{r})} t^{\mathbf{r},\mathbf{r}'} + O(\delta \mathbf{r}^2). \quad (19)$$

Above, $\delta \mathbf{r}$ is the change in distance between orbitals given by the applied (unsymmetrized) strain as $u_{\mu\nu} = \partial_\mu \delta r_\nu$ [65]. Applying this prescription to Eq. (13) we define the phonon stress tensor as

$$T_{\mu\nu}^{(p)} = \frac{\delta H(u_{\mu\nu})}{\delta u_{\mu\nu}}. \quad (20)$$

Given the structure of the viscosity tensor Eq. (5) and the fact that antisymmetric strains enter only at higher orders in $\delta \mathbf{r}$ in Eq. (19), it suffices to consider "diagonal" strains (i.e., $u_{xx}$, $u_{yy}$ and $u_{zz}$ [66]. We find that to first order (see SM [53]),

$$\begin{aligned} t_{01}, t_{23} &\to t + (u_{xx} + u_{yy})t \\ t_{02}, t_{13} &\to t + (u_{yy} + u_{zz})t \\ t_{03}, t_{12} &\to t + (u_{zz} + u_{xx})t. \end{aligned} \quad (21)$$

The diagonal phonon stress tensor restricted to the spin-1 fermions is then

$$T_{xx}^{(p)}(\mathbf{k}) = v_F \cos(\phi)(k_x L_x + k_y L_y) \\ + v_F \sin(\phi)(k_x \tilde{L}_x + k_z \tilde{L}_z) \quad (22)$$

$$T_{yy}^{(p)}(\mathbf{k}) = v_F \cos(\phi)(k_y L_y + k_z L_z) + m v_F \sin(\phi)(k_x \tilde{L}_x + k_y \tilde{L}_y). \quad (23)$$

$T_{\mu\nu}^{(p)}$ transforms as a tensor in the point group 23, which is the point group describing both the underlying lattice and the $\Gamma$ point. Note that even when $\phi = 0$, although the Hamiltonian $h$ is invariant under $SO(3)$, $T_{\mu\nu}^{(p)}$ is covariant only under the discrete group 23.

The continuity stress $T_{\mu\nu}^{(c)}$ is defined via a lattice analog of the momentum continuity equation (see SM [11]), resulting in

$$T_{\mu\nu}^{(c)}(\mathbf{k}) = \left( k_\nu \partial_\mu f(\mathbf{k}) + \frac{i}{2} \epsilon_{\mu\nu\rho} [f(\mathbf{k}), L_\rho^{\text{int}}] \right). \quad (24)$$

$T_{\mu\nu}^{(c)}$ contains contributions from "kinetic" strains (spatial deformations) and from "spin" strains due to the internal angular momentum $\mathbf{L}_{\text{int}}$. The continuity stress generalizes the Belinfante (improved) stress tensor [11,67,68].

In our model, we have $\mathbf{L}_{\text{int}}^\Gamma = 0 \oplus \mathbf{L}$ describing the spin-0 and spin-1 fermions. Using this, we can compute the stress tensor near the $\Gamma$ point restricted to the spin-1 fermion to find (see SM),

$$T_{\mu\nu}^{(c)} = \frac{v_F \cos(\phi)}{2}(k_\mu L_\nu + k_\nu L_\mu) \\ + \frac{v_F \sin(\phi)}{2} \left( 3k_\nu \tilde{L}_\mu - k_\mu \tilde{L}_\nu + \sum_{a\rho\lambda} \epsilon_{\mu\nu\rho} \Theta_{\rho\lambda}^a k_\lambda \epsilon^{ab} v_b \right), \quad (25)$$

where $a, b = 1, 2$ as in Eq. (3).

Note that $T_{\mu\nu}^{(c)} \neq T_{\mu\nu}^{(p)}$. In the continuity approach, antisymmetric stress (caused by anisotropy) enters at order $\phi$ and $T_{\mu\nu}^{(c)}$ matches the symmetries of the Bloch Hamiltonian at the $\Gamma$ point—when $\phi = 0$ the continuity stress is $SO(3)$-covariant. By contrast, when $\phi = 0$ the phonon stress is anisotropic. The distinction between the two stress tensors stems from their different physical interpretations: The phonon stress is sensitive to the nonzero orbital positions of the $4a$ Wyckoff position (see SM), which result in anisotropy in $T_{\mu\nu}^{(p)}$ when $\phi = 0$. Contrarily, the continuity stress averages over intraunit cell momentum transport and so is sensitive only to the symmetries of the effective Hamiltonian. Below, we will compute the PHV with $T_{\mu\nu}^{(p)}$ and the MHV with $T_{\mu\nu}^{(c)}$.

*Hall viscosity.* Next we compute the Hall viscosity coefficients $\eta_1$ and $\eta_2$ from Eq. (5), and the physical response $\eta_{\text{tot}} = \eta_1 + \eta_2$, for both the MHV and PHV. Focusing on the spin-1 fermion, we can simplify the Kubo formula Eq. (17) in terms of eigenstates $|n\rangle$ of $h$ as

$$\eta_{ijkl}^{\text{H}} = \frac{-1}{4\pi^3} \int d^3 k \sum_{n \neq m} \frac{\mathcal{O}_{nm}}{\Delta \epsilon_{nm}^2} \text{Im}(\langle n|T_{ij}|m\rangle\langle m|T_{kl}|n\rangle), \quad (26)$$

where $\Delta \epsilon_{nm} = \epsilon_n - \epsilon_m$, and the relative occupation factor is $\mathcal{O}_{nm} = n(\epsilon_n - \mu, T) - n(\epsilon_m - \mu, T)$ with $n(\epsilon, T) = (1 + e^{\epsilon/T})^{-1}$ the Fermi distribution with chemical potential $\mu$ and temperature $T$. We now specify to $T = 0$.

For the PHV, the stress tensor Eq. (20) is explicitly symmetric under $\mu \leftrightarrow \nu$ and therefore $\eta_2 = \eta_{1221}$ is zero. The total PHV in this case is entirely due to $\eta_1 = \eta_{1122}$ and given by (to first order in $\phi$):

$$\eta_{\text{tot}}^{(p)} = \eta_1^{(p)} = \frac{v_F^2}{8\pi^3} \begin{cases} \beta_1(-17\Lambda^3 + 60\mu^3)\phi, & \mu > 0 \\ \beta_1(-17\Lambda^3 + 42\mu^3)\phi, & \mu < 0, \end{cases} \quad (27)$$

where the momentum cutoff $\Lambda$ (Fig. 2(b)) regulates the integral in Eq. (26), and $\beta_1 = \frac{4\pi}{2835} \approx 0.00443$.

By contrast, for the MHV, the total viscosity to first order in $\phi$ is entirely due to $\eta_2$. Using the energies and eigenvectors given in the SM [53], we find that the integrand for $\eta_1$ in Eq. (26) has an energy denominator that is odd in $k_z$ at order $\phi$, which suppresses the zeroth order contribution from the numerator. When the states are taken to zeroth order in $\phi$, the numerator is odd in $k_x$ and $k_y$, and when the states are taken to first order in $\phi$, the only nonvanishing matrix elements in the numerator are odd in $k_z$, all of which leads to $\eta_1 = 0$ [69]. The total MHV is

$$\eta_{\text{tot}}^{(c)} = \eta_2^{(c)} = \frac{v_F^2}{8\pi^3} \begin{cases} \beta_2(\Lambda^3 - \mu^3)\phi & \mu > 0 \\ -\beta_2(\Lambda^3 + \mu^3)\phi & \mu < 0, \end{cases} \quad (28)$$

where $\beta_2 = \frac{4}{405}\pi \approx 0.0310$. Around $\mu = 0$, the viscosity is discontinuous. This arises from the fact that, since the antisymmetric part of the coninuity stress is linear in $\phi$, we must consider the unperturbed band structure in the energy denominators in Eq. (26). When $\phi = 0$, the band structure has a flat band bisecting two linearly dispersing bands. The filling of the flat band when $\mu$ passes through zero then causes the discontinuity in $\eta_2$, which we can attribute to the contribution of this band to the Hall viscosity. We plot $\eta_{\text{tot}}^{(c,p)}$ in Fig. 2(c).

Similar to the Hall viscosity for Dirac fermions in 2D [11,20,70], we see that both $\eta_{\text{tot}}^{(p)}$ and $\eta_{\text{tot}}^{(c)}$ consist of two terms, one of which depends explicitly on the cutoff $\Lambda$. We can interpret the cutoff-independent contribution (or, more properly, its derivative with respect to chemical potential) as the Fermi surface contribution to the Hall viscosity, while the cutoff-dependent term parametrizes unknown contributions to the viscosity from occupied states at large momenta. Using the continuity stress tensor, we can go beyond this approximation to compute the MHV for the full tight-binding model numerically (see SM).

*Conclusion.* We have highlighted a manifestly 3D cubic Hall viscosity (MHV and PHV), which appears with tetrahedral symmetry. As a proof-of-concept, we have shown that these viscosities are nonzero for a threefold fermion at the $\Gamma$ point in MSG $P2_13$ (No. 198.9). Using our phonon and continuity methods to examine the stress response in this model, we found that the MHV and PHV were nonzero in this system. We also emphasize that the MHV and PHV are responses defined for distinct stress tensors. The MHV corresponds to the "continuity" stress that exactly matches the symmetries at the $\Gamma$ point, while the PHV corresponds to the "phonon" stress which is intimately connected to the elastic response of the underlying lattice model.

Beyond our proof-of-principle calculation, the manifestly 3D nature of the cubic Hall viscosity suggests that viscous transport in 3D magnetic materials can be phenomenologically different than in two dimensions. In particular, measuring the local flow profiles [2,31] or thermoelectric transport coefficients [27] in magnets in the tetrahedral SGs (Nos. 195–206) would reveal the signatures of our 3D viscosity. For example, outside of MSG 198.9, we could consider a 3D cubic magnet with approximate Galilean symmetry at low energies. For such a system the force tensor Eq. (7), and therefore the MHV [71], is proportional to the wave-vector-dependent Hall conductivity [10,57,72],

$$\omega^2 \delta\sigma_{ij}^H \propto \eta_{\text{tot}}^{(c)} \Lambda^{mk\ell} q_k q_\ell \epsilon_{mij} \equiv \eta_{\text{tot}}^{(c)} V_\ell(\mathbf{q}) \epsilon_{\ell i j}, \quad (29)$$

where the vector $\mathbf{V}(\mathbf{q})$ highlights the structural parallel with the natural optical activity of a crystal [40]. We can decompose $\mathbf{V}$ into longitudinal and transverse components as $\mathbf{V}_\parallel = \hat{\mathbf{q}}(\hat{\mathbf{q}} \cdot \mathbf{V}) = 6\eta_{\text{tot}} q_x q_y q_z / |\mathbf{q}|$ and $\mathbf{V}_\perp = \mathbf{V} - \mathbf{V}_\parallel$. We then see that $\mathbf{V}_\parallel$ gives a $\mathbf{q}$-dependent correction to natural optical activity, while $\mathbf{V}_\perp$ leads to a Hall current proportional to the longitudinal component of the electric field. Note, crucially, that $\mathbf{V}_\parallel$ vanishes for plane waves at normal incidence. Furthermore, in tetrahedral systems without Galilean invariance, this wave-vector-dependent contribution to the conductivity need not be zero; the fate of the viscosity-conductivity relation in these systems (generalizing work such as Ref. [73]) is an interesting avenue for further study. Analogous considerations for flow in narrow channels suggest that $\eta_{\text{tot}}$ may play a role in interaction-dominated transport in narrow channels [12,13]. The physical signatures of PHV proposed in spin-phonon and electron-phonon-coupled systems [22,24–26], such as contributions to thermal Hall conductance and modifications to phononic dispersion, could be probed as well in these systems to measure the cubic PHV. The frequency and disorder [74] dependence of the MHV and PHV could also yield interesting insights. As none of our results are specific to the hydrodynamic regime, we expect disorder that preserves the symmetry on average will not modify our qualitative conclusions.

Chiral magnets such as the family of Mn$_3$IrSi [51,52] are promising platforms to study these effects. As shown in Ref. [28], this compound has a noncollinear magnetic configuration preserving the size of the unit cell; group theory analysis showed further that the ground-state magnetic order preserved all of the unitary symmetry operations consistent with MSG $P2_13$. Another interesting candidate is MnTe$_2$ in MSG $Pa\bar{3}$ (No. 205.33) [75–77]. It has a reported noncollinear magnetic structure, with the magnetic moments of the four inequivalent manganese ions pointing along the cubic body diagonals. Although naturally a semiconductor, Ag-doping could increase the carrier concentration [78].

*Acknowledgments.* P.R. and B.B. acknowledge support from the Alfred P. Sloan Foundation, and the National Science foundation under Grant No. DMR-1945058. M.G.V. and I.R. acknowledge the Spanish Ministerio de Ciencia e Innovacion (Grant No. PID2019-109905GB-C21). A.C. acknowledges financial support through European Union structural funds, the Comunidad Autonoma de Madrid (CAM) NMAT2D-CM Program (S2018-NMT-4511), and the Ramon y Cajal program through the Grant No. RYC2018-023938. A.B. acknowledges financial support from the Spanish Ministry of Science and Innovation (PID2019-105488GB-I00). The work of J.L.M. has been supported by Spanish Science Ministry Grant No. PGC2018-094626-B-C21 (MCIU/AEI/FEDER, EU) and Basque Government Grant No. IT979-16. This material is based on work supported by the National Science Foundation Graduate Research Fellowship Program under Grant No. DGE-1746047.